# A New Insight on Physical Phenomenology: a Review

**Stefano Bellucci [1,*], Fabio Cardone [2,3] and Fabio Pistella [4]**

1 INFN-Laboratori Nazionali di Frascati, Via E. Fermi 54, 00044 Frascati, Italy
2 GNFM, Istituto Nazionale di Alta Matematica "F.Severi"; Città Universitaria, P.le A.Moro 2 - 00185 Roma, Italy
3 Istituto per lo Studio dei Materiali Nanostrutturati (ISMN – CNR); c/o Università La Sapienza di Roma - 00185 Roma, Italy
4 Ente Nazionale di Ricerca e Promozione per la Standardizzazione (ENR), Italy
* Correspondence: bellucci@lnf.infn.it

**Abstract:** After a brief digression on the current landscape of theoretical physics and on some open questions pertaining to coherence with experimental results, still to be settled, it is shown that the properties of the Deformed Minkowski space lead to a plurality of potential physical phenomena that should occur, provided that the resulting formalisms can be considered as useful models for the description of some aspects of physical reality. A list is given of available experimental evidences not easy to be interpreted, at present, by means of the more established models, such as the standard model with its variants aimed at overcoming its descriptive limits; these evidences could be candidates to verify the predictions stemming from the properties of the Deformed Minkowski space. The list includes: anomalies in the double-slit-like experiments; nuclear metamorphosis; torsional antennas, as well as the physical effect of the "geometric vacuum" (as defined in analogy with quantum vacuum), in the absence of external electromagnetic field, when crossing critical thresholds of energy parameter values, energy density in space and energy density in time.

Concrete opportunities are suggested for an experimental exploration of phenomena, either already performed but still lacking a widely accepted explanation, or conceivable in application of the approach here presented, but not tackled until now. A tentative list is given with reference to experimental infrastructures already in operation, the performances of which can be expanded with limited additional resources.

## 1. Introduction

### *1.1. Multidimensional geometrical representation of physical reality*

Since the beginning of the 20th century a line of mathematical research on the multidimensional geometric formulation of physical phenomena has been active. The first to attempt this path was H. Poincaré [1] starting from the work of H. Lorentz [2] who, with his transformations, had removed the absolute nature of the time variable. These studies are part of the broader research theme related to non-Euclidean geometries that Poincaré conducted up to non-commutative geometries [3], consistent with the objective of the complete "geometrization" of the physical world according to the intentions later formulated by A. Einstein [4]. This also includes the research theme commonly called applied differential geometry, dedicated to the deepening of the properties characterizing some mathematical structures (manifold - variety - and more specifically bundles - fibers -) and the related methods of representation (in particular in the Lagrangian space, see e.g. [5] ) and of transformation; the latter in the context of physics are called gauge - calibration - , see e.g. [5, 6] and are also the subject of group theory for the research of invariants and symmetries see e.g. [7], with reference to A. Einstein's field equations. Geometrization should be regarded not as a replacement of more consolidated approaches based on forces and corresponding fields, but, in the spirit of Bohr's Principle of complementarity [8], as an auxiliary formalism whose legitimacy may derive only from experimental verification, provided any of the specific predictions that such a representation yields are verified, while those stemming from other approaches are not verified with comparable extension. An illuminating presentation of the role of theoretical physics in connection with experimental results has been formulated by M. Gell-Mann [9].

### *1.2. Potentiality of the generalization of currently used representations*

The approach based on the generalization of established representations is widely used and has often allowed an original contribution of mathematics to the advancement of knowledge on physics. As a classical example we can cite the results obtained by Minkowski [10] that made possible the representation of both electron behavior and electrodynamics in general, through a metric formulation, continuing the attempt started by Poincaré [11] and abandoned by him.

More recent examples of results obtained by generalization are the theories like string theory [12]. In turn, these proposals led to a broad mathematical development whose main result is that the maximum number of dimensions to be used in such physical-mathematical theory is fixed in a deductive way, contrary to all other theoretical models where such number is not deducted but axiomatically established by choice ad libitum. This point will be discussed more extensively in subsec. 3.3 below.

Another successful example are theories labelled "Supergravity" as commented in subsec. 3.2 below.

The adoption of the generalization tool can be considered to be of particular value when it allows an application of the *'what if'* method in a deductive path, which is both interesting in itself from a logical mathematical point of view and of potential value at the physics level, if the conjectures resulting from the generalized representation gave rise to forecasts, on a phenomenological ground, such as to be subject to experimental verification or refutation, in the wake of K. Popper's teaching [13]. It can be considered implicit in the constraint of experimental verification the selection of variables representing physical concepts in accordance with their definition by Percy Bridgman [14] who states that in physics "*any concept is nothing more than a set of operations*".

#### 1.2.1. Extension of the number of dimensions

Of particular interest for the purpose of identifying hypothetical consequences in the prediction of potential physical phenomena is the study of the generalizations that are obtained when adding a new variable to the physical variables (spatial coordinates and time) commonly used in mathematical physics.

Historically, this path was followed both by Kaluza and Klein [15], suggesting the introduction of the charge, which is a relativistic invariant, as a fifth variable, and Wesson [16], suggesting to introduce the inertia at rest, which is a relativistic invariant too, again as a fifth variable. Later, Randall and Sundrum, without advocating the compactification, proposed the introduction of a fifth variable whose physical effects, however, decay exponentially in space, see ref. [17].

Higher-dimensional generalizations of the Randall-Sundrum models with two branes and with toroidally compact subspace are considered in [18]. Z2-symmetric braneworlds of the Randall-Sundrum type, with compact dimensions, have been also considered, identifying the boundary conditions on the fermionic field, for which the contribution of the brane to the current does not vanish, when the location of the brane tends to the boundary of AdS spacetime [19]. The consideration of the Randall–Sundrum 2-brane model with extra compact dimensions allowed ~~us~~ also to estimate the effects of the hidden brane on the current density on the visible brane [20]. For a higher-dimensional version of the Randall-Sundrum 1-brane model, see [21].

In summary, as the fifth physical coordinate, extended and "measurable", both Kaluza-Klein and Wesson used a relativistic invariant together with the compactification mechanism, whereas Randall and Sundrum did not use an invariant nor compactified the just introduced additional coordinate, but have been forced to assume that its physical effects decay exponentially in space. We adopt in the present work the choice, proposed by M. Francaviglia and R. Mignani with contributions by E. Pessa [22, 23, 24, 25], to use energy as an additional fifth coordinate and also to introduce the energy dependence of all the five parameters of the metric. This gives rise to the so-called deformed Minkowski space-time-energy DM5. In the Einsteinian context of mass-energy equivalence, this choice can be interpreted as an extension of both previous proposals. Throughout the paper we refer to this approach as the MPF vision of generalized spacetime.

The results obtained in this way go beyond the limits of the choice of Kaluza and Klein, as well as of Wesson, who both failed to explain the quantization of the charge, nor succeeded in completely reconciling the representation of gravity with that of the other interactions. In turn, in the MPF proposal, both the squared charge and the Planck's constant become constants of the first integrals of the geodetic motion in DM5. At the same time, the relations, that connect energy and time in the Heisenberg's uncertainty principle, automatically stem from the conditions of the geometric motion in the 5-dimensional space-time-energy continuum (see the discussion following eq. (48) in ref. [22]). In this context there is no need to use the BRST (Becchi-Rouet-Stora-Tyutin) method.

#### 1.2.2. Extension in the use of the concept of metrics

It is important to remark that, promoting the description of gravity as a metric theory to become a method to encompass also the remaining three interactions [22, 23], yields an obvious advantage. Indeed, this approach, being general enough to include all four interactions within the same mathematical description, namely the geometric one, without forcing a unification of the interactions into a single one allows us to ultimately achieve a unification of *all* fundamental interactions at the purely geometric level.

On the other hand, focusing on possible experimental consequences of this interpretation, with reference to the contribution of metrics to the representation of physical reality, we can cite the works of M. Sachs [26] who arrived at calculating the hyperfine spectrum of hydrogen without using quantum mechanics and R. Mignani together with M. Francaviglia and E. Pessa [22, 23] who formulated predictions whose coherence with some phenomena found through experiments is the subject of this article.

### *1.3. Gauge fields as an intrinsic consequence of geometry*

Another key and empowering aspect of the MPF proposal adopted here is to extend the gauge transformation [23, 27, 28], commonly used for fields, to include the transformation of metrics. This inclusion can give a contribution to the ongoing investigations in view of the unification of forces, besides through an extension of the standard model to gravitation, also symmetrically intervening to describe strong and weak interactions, as well as the electromagnetic one, through the formalism centered on metrics, typically used for gravitational interactions.

From the modern point of view, we can advocate here that the expression "Deformed Minkowski space" (DM), regardless of the number of dimensions, should be, more significantly and perhaps more correctly, denoted as "Generalized Minkowski space", endowed with generalized Lorentz transformations, defining the conditions for the Lorentz invariance, as we shall discuss later on.

About the problem of the energy definition, we are well aware that the energy is not defined locally in general relativity. This is a consequence of the principle of equivalence, not specifically of Einstein's equations. Correspondingly, and distinctly from Newtonian gravity (where the potential can be deduced from the Schwarzschild metric, when the curvature is set to zero in the exponential form of the latter [29, 30]), in Einstein's gravity such an obstruction in defining the concept of energy persists. The MFP proposal suffers from the same difficulty. However, just as in the case of the experimental tests and validations of Einstein's gravity and General Relativity through the verifications of the phenomena based upon the notions of Riemann's curvature and torsion tensor, we wish to propose, in the spirit of Popper and Penrose, to put to the "ordeal" of the experimental test, this approach, as well as the corresponding extension in the geometry, to include also the deflection. In this sense, a characteristic signature of the MFP approach ought to emerge in the possible dependence of certain experimental phenomena from the direction in space, i.e. in the appearance of anisotropic and asymmetric features.

Historically, attention was called upon the fact that the weak and null energy conditions are violated in solutions of Einstein's theory with classical fields as material source [31]. It was shown that the discussion is only meaningful when ambiguities in the definitions of stress-energy tensor and energy density of the matter fields are resolved, with emphasis on the positivity of the energy densities and covariant conservation laws and tracing the root of the ambiguities to the energy localization problem for the gravitational field [32]. Analogously, if we consider the response of a gravitational wave detector to scalar waves in connection to the possible choices of conformal frames for scalar-tensor theories, then a correction to the geodesic equation arising in the Einstein conformal frame yields a modification to the geodesic deviation equation, thus yielding a longitudinal mode that is absent in the Jordan conformal frame [33]. On the other hand, in the MFP approach there are no problems of this kind with the energy density, because the energy densities for each interaction giving rise to new phenomena descend from invariants of the theory that are experimentally determined threshold energies separating the flat metric from the non-flat one, for each interaction. A possible further development could be the use of the stress-energy tensor as an additional variable instead of scalar energy. Presently the investigation of the possibilities opened by the extension based on the introduction of scalar energy is worthwhile to be pursued.

## 2. Features of the MPF vision of Minkowski space

*2.1 Recall on the results of the deformed Minkowski metrics*

The purpose of the present paper, starting from the MPF vision, is to show that significant results occur in the framework of the following geometric "deformation"

$$ds^2 = b_0^2(E)c^2dt^2 - b_1^2(E)dx^2 - b_2^2(E)dy^2 - b_3^2(E)dz^2 = g_{DM\mu\nu}(E)dx^\mu dx^\nu; \quad (1)$$

$$g_{DM\mu\nu}(E) = diag\left(1, -b^2(E), -b^2(E), -b^2(E)\right);$$

where $x^\mu = (ct, x, y, z)$, with $c$ denoting the vacuum light-speed.

Note that the metric tensor in (1) is diagonal; this is an important property for applications. This dynamical metric tensor yields a realization of the "Finzi Principle of Solidarity" [34]. In this sense the Solidarity Principle behaves in analogous way to the action-reaction principle. Further considerations are presented in subsec. 2.3 below.

Furthermore, we stress that, from the physical point of view, $E$ is the measured energy of the system, and thus a merely phenomenological (non-metric) variable. Here we notice, in passing, that the relevance of the Bridgman's approach to physical concepts appears. As it is well known, all the present physically realizable detectors work via their electromagnetic interaction in the usual space-time M, so $E$ is the energy of the system measured in fully Minkowskian conditions and the Hamilton theorem of total energy conservation holds. We notice in passing that, actually, owing to its very geometric nature, DM possesses nontrivial curvature, torsion and deflection, although the metric (1) describes a flat space. In the DM space, the properties of curvature, torsion, and deflection become fully manifest once one recognizes the DM space as a generalized Lagrange space. Recognition of this property is done by applying the "irreducibility theorem" (see [25, 27]), i.e., by proving that DM does not reduce to a Riemann space nor to a Finsler space, nor to an ordinary Lagrange space.

The explicit functional form of the metric (1) for all the four interactions can be found in refs. [35, 36, 37]. Here, we just wish to recall that the electromagnetic and the weak metrics show the same behaviour [32] whereas, for strong and gravitational interactions, a deformation of the time coordinate occurs (namely the time metric parameter assumes values different from unity [35, 36, 37]; moreover, the three-space is *anisotropic*, with two spatial parameters constant but different in value and the third one variable with energy in an "over-Minkowskian" way reaching the limit of Minkowskian metric for decreasing values of $E$, with $E > E_0$) [35, 36, 37].

As last remark, we stress that actually the four-dimensional energy-dependent spacetime DM is just a manifestation as a subspace of a larger, five-dimensional space in which energy plays the role of a fifth dimension. Indeed, it can be shown that the physics of the interaction lies in the curvature of such a five-dimensional spacetime, in which the four-dimensional, deformed Minkowski space is embedded. Moreover, *all* the phenomenological metrics of refs. [35, 36, 37] can be obtained as solutions of the vacuum Einstein equations in this generalized scheme [22, 23], which resembles a Kaluza-Klein like but without any compactification since the extra dimension is a true extended one: the fifth dimension is the energy $E$. We wish to recall that these metrics are part of twelve classes of solutions for the five-dimensional Einstein's equations [22, 23, 24] (see in particular Chap.20 Sect.20.3 p.296-299 of ref. [25]). For an in-depth discussion about the fifth dimension in DM5 and its phenomenological matrix, it is useful to see chapter 19, section 19.1, pp. 279-281 and note (1) at p. 280 of [25].

The generalized metric defined by eq. (1) and the corresponding interval are clearly not preserved by the usual Lorentz transformations. If $\Lambda_M$ is the 4 x 4 matrix representing a standard Lorentz transformation, this amounts to say that the similarity transformation generated by $\Lambda_M$ does not leave the deformed metric tensor $g_{DM}$ invariant:

$$(\Lambda_M)^T g_{DM} \Lambda_M \neq g_{DM} \quad (2)$$

where T denotes transpose. However, in DM it is possible to introduce Generalized (deformed) Lorentz Transformations (GLT) which are the isometries of the Deformed Minkowski space DM [19, 20]. If X denotes a column four-vector, a GLT is therefore a 4 x 4 matrix $\Lambda_{DM}$ connecting two inertial frames $K$, $K'$ such as $X' = \Lambda_{DM}(E)X$ and leaving the deformed interval invariant:

$$\Lambda_{DM}^T(E) g_{DM}(E) \Lambda_{DM}(E) = g_{DM}(E) \quad (3)$$

Therefore, unlike the case of a standard LT, a GLT generates a similarity transformation which preserves the deformed metric tensor. Let us also notice the explicit dependence of $\Lambda_{DM}$ on the energy $E$ . This means that in DM, Lorentz invariance is still valid, although in a generalized sense, namely it should be meant as the invariance with respect to the GLT. From this point of view, we can interpret the customary LT as a special transformation which finds its natural extension in the new concept of the GLT. The explicit form of the Generalized (deformed) Lorentz transformations can be found in [22, 23], whereas the geometrical aspects of the Generalized Lagrange spaces are dealt with in [38].

*2.2. Three main properties of the Generalized Lagrange space*

In ref. [27] the representation of Deformed Minkowski's space as Lagrange Generalized space is illustrated. The phenomenology of electromagnetism was systematized in a phase of physics that systematically used the schematics of the interaction of objects with a field. Consequently, a taxonomy of phenomena based on the identification of types of interaction was born. The availability of alternative readings (fields, particles, interactions, space deformations) can be considered not as an ontological difficulty, but as a richness in the light of Bohr's Principle of Complementarity. In previous works the operation that generalizes Minkowski's space has been named "deformation" and consequently the resulting space has been named Deformed Minkowski's space. In the present work, aimed at illustrating the perspectives opened by the targeted use of the term "generalization", it would have been preferable to use the broader name "Generalized Minkowski's space". Instead, the old name has been maintained for continuity with the use by the authors of previous works.

Further details on the concept of interaction and the interpretation of DM as a Generalized Lagrange space, with its irreducibility to ordinary spaces, such as Riemann space, or Finsler space, or Lagrangian space, can be found in ref. [25]. Particular attention is due to the identification of three main properties of the Generalized Lagrange space, which we consider of major importance to be investigated, if experimental results can be easily interpreted by applying the MPF vision; namely:

- the curvature defined as the application of Christoffel generalized symbols to the coefficients of the metrical canonical $H$−connection of generalized Lagrange space $GL^N$;
- the torsion defined as the distinguished tensors, d-tensors, of the metrical connection of the generalized Lagrange space $GL^N$;
- the deflection defined as the horizontal (**h**) and vertical (**v**) covariant derivatives of the Liouville vector fields on the tangent manifold (TM), defined in eq. (35) of ref. [27].

*2.3 Internal fields as a way to "store" the energy of the deformation*

As we have seen, internal fields store the energy of the deformation and return it in the form of the effect of the deformation, i.e. modification of the measurement of time and therefore different release of energy by space-time, see eq. (52) of ref. [27], compared to that of the interaction, see eq. (37) of ref. [27]. This, for example, can be a new key to reading phenomena in the field of elementary particles such as "asymptotic freedom" and "confinement". In phenomenological terms, eqs. (37), (52) of ref. [27] can tell us that, with re-spect to an electromagnetic stress, asymptotic freedom is a "slow" response while confinement is a "fast" response. In simple terms, the principle of action and reaction depends on the measure of time to have a response with respect to the measure of time relative to a stress. This effect entirely replaces the idea that there is a temporal inertia, i.e. a latency time that separates the action from the reaction. Moreover, as mentioned above, the eqs.(16), (37), (50-52) of ref. [27] constitute the basis of hypotheses of the "Mignani mimicry" to explain also nuclear reactions without exceeding the Minkowskian threshold of energy $E_0$ for the metric of nuclear interactions. This can happen when the electromagnetic field inside the electromagnetic space itself, in absence of an external electromagnetic field, "mimics", thanks to the eq. (52) of ref. [27], a property of the nuclear space. In this sense there can also be an alteration of the reproducibility of phenomena if these conditions are not taken into account (see also sec. 4).

## 3. Current theoretical landscape

### *3.1. The maximal symmetry of the S-matrix and the arising of supersymmetry*

So far, theoreticians attempted to create a unified mathematical structure, complemented with *ad hoc* physical assumptions, in ordinary Minkowski spacetime, capable of encompassing, within the same framework, gauge theories providing the description of three of the four fundamental interactions related to the phenomena of electromagnetism, weak and strong nuclear reactions - however without any metric representation -, together with a quite different interaction, i.e. gravity, which is properly described by a metric theory. In this section, we give a brief description of the current state of the art of the available approaches in such topics, that are built upon alternative kinds of generalizations, as compared to the one put forward in this work.

First of all, it is worth recalling here a well-known result, originally proved in the second half of the 20th century by Coleman and Mandula [39]. Coleman and Mandula addressed the fundamental question, about the maximal symmetry allowable for the scattering matrix. They came, under five quite general and natural assumptions - e.g. Lorentz invariance, particle finiteness, elastic analyticity (in weak form), the existence of scattering -, to formulate the celebrated "no-go theorem" stating, that the S-matrix allows, as its most general symmetry group, the direct product of the Poincaré group and an internal symmetry group. More recently [40], Weinberg gave an alternative proof of the theorem, which avoids group considerations and focuses on the symmetry (Lie) algebras. In other words, rather than working with the symmetry group, he worked with the algebra of the corresponding generators. Of course, returning to groups consideration involves, then, just a simple further step.

In spite of the generality of its assumptions, the theorem is unable to capture all possible occurrences. For instance, in massless theories additional generators of the conformal group, mixing with the Poincaré group generators, can occur. Also, both cases of spontaneously broken symmetries, as well as discrete symmetries escape the theorem, since they do not commute with the S-matrix. Furthermore, the theorem does not apply to 1+1 dimensional theories, where scattering angles cannot obey the assumed analyticity requirement, given that, in such theories, only forward and backward scattering can occur (see [41]).

Probably the most famous class of theories evading the theorem are supersymmetric theories, as the implicit assumption made by Coleman and Mandula was that internal symmetries generators had to necessarily combine trivially with the generators of the space-time symmetries, i.e. by commuting with them. It was the consideration of graded Lie algebras that gave rise to the saga of 4-dimensional supersymmetric theories, all of them violating, so to speak, the famous theorem, and hence, at least at the beginning, quite unexpected. In 1971 Yu. Golfand and E. Likhtman [42] generalized the Coleman–Mandula theorem to include the case of supersymmetry without central charges generators. It was in the 1975 paper by Haag–Łopuszański–Sohnius where finally the theorem was generalized, to show that the possible symmetries of a consistent 4-dimensional quantum field theory can also include supersymmetry with central charges, as a nontrivial extension of the Poincaré algebra [43]. The first consequence of the theorem including supersymmetry generators was the ruling out of the spin higher than 1/2 from the fermionic part of the Lie superalgebra. The second was to provide a justification for a result published one year before by Wess and Zumino who proposed a renormalizable, interacting 4-dimensional quantum field theory which contained linearly realized supersymmetry [44].

*3.2 Supergravity theories, as candidates for a unified description of general relativity and quantum field theory*

In 1976 Freedman, Ferrara and van Nieuwenhuizen [45] set up the minimal version of 4-dimensional supergravity, as an example of field theory combining supersymmetry with general relativity. Shortly afterwards, Deser and Zumino [46], independently proposed the minimal 4-dimensional supergravity theory. For a time, it was hoped that such a theory would allow handling an unresolved issue in theoretical physics, i.e. the harmonization of quantum field theory, describing the three fundamental interactions among elementary particles, with general relativity.

The problem is well known: the unavoidable occurrence of infinities arises when considering the theories beyond the classical approximation, to include radiative quantum corrections. While, on the one side, for the three fundamental interactions of electromagnetism, weak and strong nuclear ones, the renormalization procedure allows to reabsorb such infinities in a finite number of parameters of the theory, on the other hand gravity escapes such a treatment [47]. As we know since 1974 [48], pure gravity theory in four dimensions is one-loop finite, although this property does not extend to the theory in the presence of matter, which is divergent already at one loop. At two loops, the pure gravity counterterm constructed from the tensor product of three Riemann tensors has a non-zero coefficient, as shown by an explicit calculation [49], see also [50]. For any supergravity theory, with any number of supersymmetric charges, however, it turns out that this tensorial product of three Riemann tensors is not a valid supersymmetric counterterm, i.e. the appearance of such a structure would produce an helicity amplitude incompatible with supersymmetry [51, 52]. Hence, in supergravity, the first divergent counterterm can show up at orders higher than two, i.e. from three loops onward. Indeed, the square of the Bel-Robinson tensor is expected to show up as a possible counterterm, admitting a supersymmetric extension, just at the three loop-level in pure, 4-dimensional supergravity [53, 54, 55]. In passing, we note that it has become by now a classic computation, that of the one-loop quantum General Relativity contribution to the anomalous magnetic moment of a lepton [56]. As explained in [57], the ultraviolet finiteness of the net result occurs, in spite of the ultraviolet divergence of each Feynman diagram contribution, as expected by the observation that each Feynman diagram is not ultraviolet finite by power counting.

*3.3 The advent of superstrings and the additional dimension*

Recently the common wisdom, according to which it is to be expected that all supergravity theories will diverge at three loops, has been reconsidered [58, 59, 60, 61], for various reasons, especially because any loop calculation carried out so far ascertained that the power counting of N = 8 supergravity yields the same result as N = 4 super-Yang-Mills theory, which is an ultraviolet finite theory. Although N = 8 supergravity might ultimately turn out to provide an ultraviolet finite gravity theory (note that some of the counterterms cancellations observed in N = 8 supergravity appear not to originate from the requirement of supersymmetry invariance, adding a further mystery to the issue), however, since the 1970s, many people have partly turned away from point-like theories, in the quest for a consistent description of gravity as a quantum theory. This attitude led to the advent of superstrings, a theory which aims at resolving the issue of making general relativity compatible with quantum mechanics, by replacing the classical idea of point particles with strings, which have an average diameter of the Planck length, i.e. the distance travelled by light in the vacuum during one unit of Planck time. Mathematical consistency of string theories requires the presence of additional spacetime dimensions, over the customary 3+1 dimensions. While the bosonic string lives in a 26-dimensional spacetime, superstrings require 10 dimensions (3 dimensions for ordinary space + 1 time dimension + a 6 dimensional hyperspace) [62].

There are two ways to reconcile, in the conventional wisdom, the lack of phenomenological evidence for the extra dimensions. The first mechanism consists in assuming that elementary particles live only on a 3-dimensional submanifold corresponding to a brane, whereas gravity does not. Alternatively, one may hypothesize the compactification of the additional dimensions on an extremely small scale. However, far from being a natural, or at least plausible feature, the mechanism of compactification of additional dimensions has been pinpointed by Richard Feynman as a drawback of superstring theories altogether. To put it in his words: "For example, the theory requires ten dimensions. Well, maybe there's a way of wrapping up six of the dimensions. Yes, that's possible mathematically, but why not seven? When they write their equation, the equation should decide how many of these things get wrapped up, not the desire to agree with experiment. In other words, there's no reason whatsoever in superstring theory that it isn't eight of the ten dimensions that get wrapped up and that the result is only two dimensions, which would be completely in disagreement with experience." [63]. That would be an authoritative opinion leaning towards considering seriously a possible, or even unavoidable additional dimension – as it seems to occur in the theory expressed in the present paper – as a verifiable property of nature, with - at least in principle - phenomenologically detectable consequences. Or, to put it in the words of another influential theoretical physicist: "So far, string theory fails to describe our world as see it. It describes, instead, lots of worlds, in all sort of higher dimensions, generally with cosmological constant having the wrong sign, with "microscopical" internal spaces of cosmological size, and so on. This is a beautiful theoretical world, with marvels and surprises, but where is our world in it? Until the description of our world is found in this immense paper edifice, it seems to me that caution should be maintained." [64].

A fascinating exception, so far never embedded into string theory, may be the DGP model of gravity, which assumes an action consisting of the customary 4-dimensional Einstein–Hilbert action - dominating at short distances -, along with an additional term equivalent to the 5-dimensional extension of the Einstein-Hilbert action, which dominates at long distances [65]. The interest of this theory lies not only in its aspiration to reproduce the cosmic acceleration of dark energy without any need for a small but non-zero vacuum energy density, but also in the claim that the unusual structure of the graviton propagator makes non-perturbative effects important in a seemingly linear regime, such as the solar system [59].

In concluding this section, let us mention, in passing, a particularly interesting case of string theory, i.e. that in one dimension, studied originally starting from a discretized approximation [66]. It is possible to show that the sum over the surfaces with different genus can be carried out for the vacuum energy yielding a result that can be written in a closed form [66]. Building upon this model of a discretized string, one can then consider its supersymmetric extension. The full string theory is defined in [61] as the sum over the triangulations of the surface - embedded in the superspace – which is described by the Wess-Zumino model, in the continuum limit. The Feynman graph expansion for the model generates a discretization of random surfaces in superspace. In this supersymmetric 1-dimensional string it turns out, by explicit computation, that supersymmetry is spontaneously broken [67]. Related work on supersymmetric matrix models includes [68].

*3.4 Central questions still pending in the physics of fundamental interactions*

These theories, which leave unaltered the Minkowski space, so far have not been able to fully cover unresolved central questions still pending in physics such as:

1. combining general relativity and quantum theory into a single theory that can claim to be the complete theory of nature, i.e. the problem of quantum gravity;

2. resolving the problems in the foundations of quantum mechanics, either by making sense of the theory as it stands or by inventing a new theory that does make sense;

3. determining whether or not the various particles and forces can be unified in a theory that explains them all as manifestations of a single, fundamental entity;

4. explaining how the values of the free constants in the standard model of particle physics are chosen in nature;

5. explaining the possible missing matter and the dark energy, as well as the possible modifications of gravity on large scales.

Among them, the MPF approach can give a contribution to address at least the first three questions, as well as the last one.

In addition, there are several perhaps not yet fully established experiments, yielding results unexplained by the current theoretical wisdom, that may find a suitable explanation within the MPF approach. In the next section we discuss both the latter experimental results and their corresponding interpretation, within the MPF framework.

## 4. Possible experimental phenomena expected from the properties of Deformed Minkowski Space and available candidate evidences

### *4.1. Anomalies in the double-slit-like experiments*

Striking evidences of unconventional photon behavior have been collected in many experimental instances [69, 60, 71, 72]. We stress that the anomaly occurs regardless of the photon frequency range, as demonstrated by the experiments carried out in the microwave and the visible range. The experimental set-up employed in the experiments directed by R. Scrimaglio [69, 70, 71, 72] showed invariably that photons anomalously, even as similar experiments employed a varied setup of detectors and photon sources. Indeed, the same anomalous result was always obtained [69, 70, 71, 72]. Carrying out a separate class of laser-based experimental activities in the microwave frequency range [73, 74], provided an independent evidence of the anomaly, which, in the DM perspective, is quite straightforward to be interpreted as a consequence of the space-time deformation, called the "shadow of light " [27].

Elaborating the interpretation more in detail we recall that the structure of the deformed Minkowski space DM as Generalized Lagrange Space implies the presence of two internal electromagnetic fields, the horizontal field $F_{\mu\nu}$ see eq. (53) of ref. [27] and the vertical one, $f_{\mu\nu}$ see eq. (54) of ref. [27]. Whereas $F_{\mu\nu}$ is strictly related to the presence of the external electromagnetic field $F_{\mu\nu}$ , vanishing if $F_{\mu\nu} = 0$, the vertical field $f_{\mu\nu}$ is geometrical in nature, depending only on the deformed metric tensor $g_{DM\mu\nu}(E)$ of $GL^4$ = DM and on the variable appearing in the metric (1) $E$, see eq. (54) of ref. [27]. Therefore, it is present also in space-time regions where no external electromagnetic field occurs eqs. (54-58) of ref. [27]. In our opinion, the arising of the internal electromagnetic fields associated to the deformed metric of DM, as a Generalized Lagrange space, is at the very core of the physical, dynamic interpretation of the experimental results on the anomalous photon behavior. Namely, the dynamic effects of the hollow wave of the photon, associated to the deformation of space-time, which manifest themselves in the photon behavior contradicting both classical and quantum electrodynamics, arise from the presence of the internal v-electromagnetic field $f_{\mu\nu}$ (in turn strictly connected to the geometrical structure of DM).

Moreover, as already stated at the end of the previous section, in the framework of DM, the role of the quantum vacuum is played by the geometric vacuum, whose physical effects are represented by the vertical field $f_{\mu\nu}$ which is responsible for the "shadow of light" effect. Actually, this effect can also be interpreted as an example of the effective action of the "geometric vacuum" which is "full" of deformation due to the energy stored in the geometry of Deformed Minkowski space in the Generalized Lagrange space in the sense explained in subsec. 1.3.

### *4.2 Nuclear Metamorphosis*

The geometrical concepts of torsion and deflection lead us naturally towards anisotropic and asymmetric phenomena, respectively for the first and the second concepts. The foundation for such predictions and the corresponding observations, lies in the "*principle of reinterpretation*" of the physical laws, due to R. Mignani et alii [75], in coherence with similar work carried out by U. Bartocci as reported in [76]. Experimental observations al-lowed to study emissions of $\alpha$-particles, as well as neutrons, that confirm those phenomena, with the predicted characteristics [77, 78]. Specifically, the fact that an angle of asymmetry is present when the system is invariant under generalized Lorentz transformations, manifesting itself through the appearance of a preferred direction of emission for $\alpha$-rays and neutrons, as a result of the spacetime deformation, can be advocated recalling the earlier work in [79]. Furthermore, the measurement of the neutron spectra produced in the condition of Deformed spacetime, illustrated in [78], which have new and unique characteristics, differing from standard neutron sources emission, can also be interpreted as a consequence of the invariance of the system under generalized Lorentz transformations.

The so-called "nuclear metamorphosis" is a phenomenon, in which materials subjected to sonication (i.e. exposure to ultrasounds) produce elements of both greater and lesser atomic weight than the elements contained in the material before sonication. This phenomenon has been observed both in steel rods (AISI 304 alloy) [79, 80], and in high purity mercury (purity level above 99% for use as a chemical reference sample) [81, 82]. Initially this phenomenon generated speculations on nuclear metabarism [83, 84]. But later the evidence of neutron emissions, both asynchronous in time and anisotropic and asymmetric in space and also anomalous in spectrum and fluence in energy [85], produced by sonicated steel with ultrasounds in the same conditions of "nuclear metamorphosis" have directed the reflection on the non-flat nature of nuclear space-time [86].

Results compatible with the present interpretative framework can be obtained by the compression of rocks [87]. Although the description and interpretation of the experimental data does not lead to firm and ultimate conclusions, the data of the neutron emission were measured by different kinds of detectors, which supports their trustworthiness. We stress that in the experimental setup aiming to test anisotropic and/or asymmetric phenomena, the detectors positions cannot be chosen arbitrarily, as they must follow the geometry of the deflection and the torsion [85]. According to this prescription, in measuring neutrons stemming from rock fracture [87] carried out according to Deformed Minkowski Spacetime neutron detectors, gave non vanishing results only when positioned according to deflection and torsion.

Finally, by applying the same ultrasonic sonication technique to substances containing a radionuclide, a significant reduction in activity has been found and analyses have confirmed that this reduction is due to the nuclear metamorphosis of the radionuclide into other inactive nuclides under matter conservation conditions [88, 89, 90].

Indeed, all these phenomena, summarized under the label "nuclear metamorphosis" [79, 80, 81, 82, 85, 88, 89, 90], always occurred in the absence of any kind of gamma emissions above the natural background, thus excluding that they could be interpreted as phenomena connected to customary nuclear transformations. Hence, this consideration yields the possibility to interpret this absence of emitted energy in the sense of Mignani's mimicry. In other words, such missing energy is the one described by the eq. (6), while the energy received by the material system corresponds to the one described in the eq. (37) of ref. [27], giving rise to a "geometric mimesis" in the sense described by the eqs. (50-52) of ref. [27].

More in general, in the spirit of the authors of ref. [86], there are indications that connect the "nuclear metamorphosis", as well as the emission of neutrons and $\alpha$-particles that accompanies them (anomalous under every aspect, temporal, spatial, energetic), to both the curvature and the deflection tensor of Deformed Minkowski space, understood as Lagrange Generalized space, see sec. 2.

Finally, it is tantalizing to also attempt an interpretation of the rich phenomenology collected during decades of experiments on low energy nuclear reactions (LENR), although they are still lacking a conclusive understanding, in the light of the theoretical concepts and ideas, on which our interpretation of the nuclear metamorphosis experiments is based. For a comprehensive review of such phenomenology, see e.g. [91]. Additional experimental results are expected from the outcomes of the H2020-EIC-FETPROACT-2019 project "Clean Energy from Hydrogen-Metal Systems (CleanHME)" project, which is cur-rently underway. A possible strategy for the theoretical interpretation of LENR experimental results would be that of searching for hints of structural modifications, critical energy density and mechanisms of releasing the loaded energy in a suitable interval of time that can be possibly explained and modeled, in the spirit of the Mignani mimicry, through the possible use of geometric concepts, such as curvature, torsion and deflection.

*4.3 Torsional antennas*

Finally, we want to mention the case of torsional antennas. It is well known that emission diagrams of antennas have peculiarities that are commonly interpreted in terms of angular energy distribution. Let us consider the case of torsional antennas composed of two concentric circular dipoles on orthogonal planes which are subjected to torsion according to an axis passing through the common center and given by the intersection of the two orthogonal planes each containing one of the circles. These torsional antennas generate the phenomenon of the symmetrical emission of electromagnetic waves, with circular polar diagram, but only for torsion angles higher than the flat angle, while, for specular angles, the polar diagrams are not comparable among themselves [92, 93]. Until now, it has not been possible to explain these phenomena with simulations based on the classical equations of the electromagnetic field. The behavior of the torsional antenna emissions seems to follow the features of the "torsion" in the Generalized Lagrange space of Deformed Minkowski as suggested by eq. (32) of ref. [27]. Also experiments conducted on torsional antennas tend to confirm the considerations exposed in Sec. 2.1 dealing with situations where the usual Lorentz transformations are not kept, while Generalized (deformed) Lorentz Transformations are maintained. The effects of twisting a double loop torsional antenna have been addressed recently in refs. [92, 93].

*4.4 Suggestion for further analysis and experimental activity*

Having discussed some results showing anomalous effects, which could be explained through a quite straightforward interpretation in terms of the intrinsic gauge fields of Deformed Minkowski space, we stress that, in our opinion, a further verification of such findings, using all needed experimental precautions (see subsec. 4.2), would be highly desirable, also in connection with the emphasis on falsifiability and the value of falsification activities included in Popper's epistemological positions.

Moreover, the deflection property that characterizes the MFP theoretical framework might be useful also in analyzing the Cosmic Microwave Background radiation data [94], in order to explore their anisotropy and asymmetries.

## 5. Concluding remarks and outlook on experimental opportunities

In the MPF approach three main properties emerge, namely:

- the curvature property is concordantly related to gravitational interaction;
- the torsion property is related to the phenomenon of anisotropy (asymmetric angular behavior) found in nature, as in the case of the torsional antenna, and anisotropic neutron emissions;
- the deflection property is connectable to the asymmetry phenomenon found in nature as in the case of cosmic microwave frequency background radiation, to the violation of parity symmetry in the lepton interaction and the asymmetric emission of neutrons and alfa particles.

The latter properties proved valuable in attempting a consistent interpretation of several otherwise unexplained phenomena, see sec. 4. We now suggest a list of possible experimental activities, aiming at verifying whether new phenomena connected with the DM Generalized Lagrangian Space properties, can really be found. It is useful to deepen the study of the properties of the torsion in DM, also in the characteristics of the gravitational interaction that result from the astrophysical evidence. Also, it is convenient to consider the possibility to handle the appearance of anisotropic features in certain experimental phenomena both at laboratory scale and at cosmological scale, using jointly torsion and deflection.

The measurement by the UA1 collaboration of the Bose Einstein Correlation in proton anti-proton annihilation in the ramping runs in 1986-87 opened new perspectives for the experimental determination of the features of the hadronic metrics [35, 95, 96]. On the one hand, there is still the unclear problem of why the value of Bose Einstein's correlation constant is anomalous, i.e. it is not the one predicted by the theory of the standard model and the quark model. On the other hand, an in-depth study of the correlation allows to improve the knowledge of the hadronic metric that was obtained for the first time with this method [35, 95]. In particular, the ramping run from 35 GeV/c up to 350 GeV/c in the center of mass was a precursor of the experimental program that ought to be carried out systematically in future; one does not have to make a single-energy measurement, i.e. carry out an experimental determination with fixed energy, but a measurement in a wide energy range, because only in this way it will be possible to determine the dependence of the hadronic metric on energy, thus opening the way to a new season of phenomenology in the physics of elementary particles.

This program can be carried out at CERN through the LHC accelerator (which has the detectors already available, i.e. those related to the quest for the Higgs particle, namely ATLAS and CMS), Tevatron Fermilab (with detectors to be searched, e.g. KLOE),

An experimental, small-scale program can also be carried out at KEK, the old LEP accelerator, or even with DAPHNE. In fact, the use of positron electron beams makes the experiment much cleaner, because little affected by the phenomena of hadronization, through the study of the inverse Compton radiation that occurs in the impact of a charged particles beam accelerated with the thermal radiation of the same vacuum pipe of the accelerator machine to measure the "shadow of light" effect due to the geometric field $f_{\mu\nu}$ and the geometric vacuum, as it has been described in the present work.

Hence, the inverse Compton radiation becomes the probe with which to verify the effect of the geometric electromagnetic field, in the absence of an external electromagnetic field, and therefore the effect of the geometric vacuum that we have presented here as a possible alternative to quantum vacuum. Here a further development of detector technology is required for the study of the gamma radiation produced by the inverse Compton effect, since both the spatial and temporal distribution (to be obtained through the study of the concomitance between beam and detection) and the energetic distribution should be studied, as was done with the "spaghetti" gamma calorimeter, calibrated at the LEP accelerator of CERN, using inverse Compton radiation in the period 1992-4 [97].

Investigating five-dimensional theories is appealing from many viewpoints, some of which we have glanced upon in subsec. 1.2.1. It is remarkable that supersymmetric black holes may live in no more than five dimensions. An example is given by the study of the attractor mechanism carried out in ref. [98], within the framework of five dimensional Einstein-Maxwell Chern-Simons theory. We recall that, compactifying M-theory on a CY3 manifold yields a five-dimensional supergravity theory, which represents the low energy approximation of the full theory. Yet, in spite of its relatively simple formulation, the effective, approximate theory grasps many of the properties of the complete M-theory.

In turn, the use of the term "interaction" is a legacy of the pre-Einsteinian phase of the development of physics, during which the phenomenology of gravity was explained in terms of interaction between masses rather than a property of space-time. There are interesting attempts to describe the mechanical and quantum-mechanical properties of the empty space-time, e.g. Rovelli [99], as well as its thermodynamical properties in terms of "atoms" of space-time Padmanbhan [100]. In particular, the latter have been introduced to try to justify the use of thermodynamical concepts, such as the entropy of black holes, for example, from a microscopic point of view. In the present work, we prefer to stick to a customary view of the space-time as a geometrical nontrivial (nonflat, or rather not strictly pseudo-Euclidean) notion.

**Acknowledgments:** We wish to thank Massimo Porrati for useful comments on the definition of energy and its meaning in General Relativity and Gravitation. We also thank Sergio Benenti for helpful hints concerning the meaning of generalized Lagrange space for DM.

## References

[1] H. Poincaré, *La théorie de Lorentz et le principe de réaction, Archives néerlandaises des sciences exactes et naturelles* vol.5 p. 252 (1900). Reprinted in Poincaré, Oeuvres, tome IX , p.464.
[2] H. A. Lorentz, *Versuch einer Theorie der elektrischen und optischen Erscheinungen in bewegten Körpern* (ed. E.J. Brill, Leiden, 1895).
[3] A. Connes, M. Marcolli, *Non commutative Geometry Quantum Fields and Motives,* American Mathematical Society ed. Colloquium Pub. Vol. 55 (2007).
[4] A. Einstein, Conference *'Geometrie und Erfahrung'* (*Geometry and Experience*), 1921.
[5] N. Steenrod, *The Topology of Fibre Bundles* (Princeton Univ. Press, 1951).
[6] R. Miron, A. Jannussis and G. Zet, in *Proc. Conf. Applied Differential Geometry - Gen. Rel. and The Workshop on Global Analysis, Differential Geometry and Lie Algebra*, 2001., Gr. Tangas ed. (Geometry Balkan Press, 2004), p.101.
[7] E. Wigner, *Group Theory* (Academic Press, New York, 1959); D. Lichtenberg: *Unitary Symmetry and Elementary Particles* (Academic Press, New York, 1978 2nd ed.).
[8] M. Beller *The Genesis of Bohr's Complementarity Principle and the Bohr-Heisenberg Dialogue*. In: Ullmann-Margalit E. (eds) *The Scientific Enterprise. Boston Studies in the Philosophy of Science*, vol 146. (Springer, Dordrecht 1992)
[9] M.Gell-Mann, *The Quark and the Jaguar*, (W.H. Freeman and Co., New York 1994) p. 89)
[10] H. Minkowski, *Raum und Zeit (Space and Time)* Physikalische Zeitschriff 10, 75 (1908).
[11] H. Poincaré, *Sur la dynamique de l'électron*, in Rendiconti del Circolo Matematico di Palermo vol.21 p.129 (1906). Reprinted in Poincaré, Oeuvres, tome IX , p.494.
[12] J. Polchinski, *String Theory vol.1,2* , Cambridge Monograph on Mathematical Physics (Cambridge, 2005).
[13] K. Popper, *Conjecture and Refutation,* Routledge Classic vol.17 , 1st ed. 1963;
R. Penrose, *The emperor's new mind*, Oxford University Press (1989)
[14] P. W. Bridgman, *The Logic of Modern Physics*, (New York: Macmillan 1927).
[15] T. Kaluza, *Sitzungsber Preuss. Akad. Wiss, Berlin. (Math. Phys.)*, 966-972, (1921). O. Klein, *Z. Phys.*, 37, 895-906 (1926).
[16] P. Wesson, *Space Time Matter* (World Scientific, Singapore, 1999).
[17] L. Randall and R. Sundrum, Phys. Rev. Lett. 83, 3370 (1999); 83, 4690 (1999).
[18] S. Bellucci, A. A. Saharian, H. G. Sargsyan, V. V. Vardanyan, *Physical Review D* 101 (4), 045020 (2020).
[19] S. Bellucci, A. A. Saharian, D. H. Simonyan, V. V. Vardanyan, *Physical Review D* 98 (8), 085020 (2018).
[20] S. Bellucci, A. A. Saharian, V. Vardanyan, *Physical Review D* 93 (8), 084011 (2016).
[21] S. Bellucci, A. A. Saharian, V. Vardanyan, *Journal of High Energy Physics* 2015 (11), 92 (2015).
[22] M. Francaviglia, R. Mignani, F. Cardone, *General Relativity and Gravitation* 31, 7, 1049 (1999).
[23] M. Francaviglia, R. Mignani, F. Cardone, *Foundation of Physics Letters* 12, 281 and 347 (1999);
R. Mignani, E. Pessa, G. Resconi, *Physics Essays* 12, 61-79 (1999).
[24] F. Cardone, R. Mignani, *Energy and Geometry - An Introduction to Deformed Special Relativity* (World Scientific, Singapore, 2004).
[25] F. Cardone, R. Mignani, *Deformed Spacetime - Geometrizing Interactions in Four and Five Dimensions* (Springer, Heidelberg, Dordrecht, 2007).
[26] S. Sachs, *General Relativity and Matter* (Reidel Pub. Co. , Dordrecht, 1982); *Quantum Mechanics from General Relativity* (Reidel Pub. Co. , Dordrecht, 1986).
[27] R. Mignani, F. Cardone, A. Petrucci, *Electronic J. of Theoretical Phys*. 10, 29, 1-20 (2013).
[28] R. Mignani, F. Cardone, A. Petrucci, *Natural Science, Special Issue* 6, 339-410 (2014) DOI:10.4236/ns.2014.66040.
[29] L. D. Landau, E.M. Lifshitz, *The Classical Theory of Fields*, Pergamon Press (1971).
[30] C. W. Misner, K. S. Thorne, J.A. Wheeler, *Gravitation*, Princeton University Press (2017).
[31] C. Barcelo and M. Visser, *Classical and Quantum Gravity*, 17, 3843 (2000).
[32] S. Bellucci, V. Faraoni, *Nuclear Physics B* 640, 453-468 (2002).
[33] S. Bellucci, V. Faraoni, D. Babusci, *Physics Letters* A 282, 357-361 (2001).
[34] B. Finzi: "Relatività Generale e Teorie Unitarie", in Cinquant'anni di Relatività (in Italian), ed. M. Pantaleo (Giunti, Firenze, Italy, 1955), pg. 194.;
F. Cardone, R. Mignani, A. Petrucci, *"The principle of solidarity: Geometrizing interactions*", in *Einstein and Hilbert: Dark matter*, V. V. Dvoeglazov ed. (Nova Science, Commack, NewYork, 2011), p. 19.
[35] F. Cardone, R. Mignani, *Journal of experimental and theoretical physics* 83, 435 (1996).
[36] F. Cardone, R. Mignani, *Gravit. Cosmol.* 4, 311 (1998).
[37] F. Cardone, R. Mignani, *International Journal of Modern Physics A*, Vol. 14, No. 24 3799-3811 (1999).
[38] R. Miron and M. Anastasiei, *The Geometry of Lagrange Spaces: Theory and Applications* (Kluwer, 1994);
R. Miron, D. Hrimiuc, H. Shimada and S. V. Sabau, *The Geometry of Hamilton and Lagrange Spaces* (Kluwer, 2002).
[39] S. Coleman, J. Mandula, *Phys. Rev.* 159, 1251 (1967).
[40] S. Weinberg: *The quantum theory of fields*, Vol. 3, (Cambridge University Press 2000).
[41] E. Witten, in: *The unity of fundamental interactions. Proceedings, International School of Subnuclear Physics, Erice 1981*, ed. A. Zichichi (Plenum Press,1983), p.305.
[41] Yu. A. Golfand, E. P. Likhtman, *JETP Lett.* 13, 323 (1971).
[43] R. Haag, M. Sohnius, J.T. Łopuszański, *Nuclear Physics B* 88, 257 (1975).
[44] J. Wess, B. Zumino, *Nuclear Physics B* 70, 39 (1974).
[45] D. Z. Freedman, P. van Nieuwenhuizen, S. Ferrara, *Physical Review D* 13, 3214 (1976).

[46] S. Deser, B. Zumino, *Physics Letters B* 62, 335 (1976).
[47] J. Polchinski, *String Theory: Volume I*, (Cambridge University Press 1998), p. 4.
[48] G. t'Hooft, M. Veltman, *Annales de l'I. H. P., section A*, tome 20, no 1 (1974), p. 69.
[49] M. H. Goroff, A. Sagnotti, *Nuclear Physics B* 266, 70 (1986).
[50] A. E. M. van de Ven, *Nuclear Physics B* 378, 309 (1992).
[51] M.T. Grisaru, *Physics Letters* 66B, 75 (1977).
[52] E.T. Tomboulis, *Physics Letters* 67B, 417 (1977).
[53] S. Deser, J. H. Kay, K. S. Stelle, *Physical Review Letters* 38, 527 (1977).
[54] M. Kaku, P. K. Townsend, P. van Nieuwenhuizen, *Physical Review Letters* 39, 1109 (1977).
[55] S. Ferrara, B. Zumino, *Nuclear Physics B* 134, 301 (1978).
[56] C.P. Martín, *Journal of Cosmology and Astroparticle Physics* 2017 (2017).
[57] S. Bellucci, H.-Y. Cheng and S. Deser, *Nuclear Physics B* 252, 389 (1985).
[58] M.B. Green, J.H. Schwarz, *Nuclear Physics B* 198, 441 (1982).
[59] Z. Bern, L. Dixon, D.C. Dunbar, M. Perelstein, J.S. Rozowsky, *Nuclear Physics B* 530, 401 (1998).
[60] N. E. J. Bjerrum-Bohr, D.C. Dunbar, H. Ita, W.B. Perkins, K. Risager, *JHEP* 0601, 009 (2006),
[61] Z. Bern, J.J. Carrasco, L.J. Dixon, H. Johansson, D.A. Kosower, R. Roiban, *Physical Review Letters* 98, 161303 (2007).
[62] J. H. Schwarz, *Nuclear Physics*, B46, 61 (1972).
[63] R. Feynman, 1987 interview published *in Superstrings: A Theory of Everything?* (1988) edited by Paul C. W. Davies and Julian R. Brown, p. 193-194 ISBN 0521354625.
[64] C. Rovelli, Foundations of Physics, 43, 8 (2011).
[65] G. Dvali, G. Gabadadze, M. Porrati *Physics Letters B* 485, 208 (2000).
[66] G. Parisi, *Physics Letters B* 238, 209 (1990).
[67] E. Marinari, G. Parisi, *Physics Letters B* 240, 375 (1990).
[68] S. Bellucci, *Physics Letters* B 257 (1-2), 35-39 (1991);
S. Bellucci, *Nuclear Physics B-Proceedings Supplements* 20, 737-740 (1991);
S. Bellucci, T.R. Govindrajan, A. Kumar, R.N. Oerter, *Physics Letters B* 249 (1), 49-55 (1990).
[69] F. Cardone, R. Mignani, W. Perconti, R. Scrimaglio, *Phys. Lett, A* 326, 1 (2004).
[70] R. Scrimaglio et al., *Int. J. Mod. Phys. B* 20,1, 85 (2006).
[71] R. Scrimaglio et al., *lnt. J. Mod. Phys. B* 20,9, 1107 (2006).
[72] R. Scrimaglio et al., *Annales Fond. L. de Broglie* 33, 319, (2008).
[73] A. Ranfagni, D. Mugnai, R. Ruggeri, *Phys. Rev. E* 69, 027601 (2004).
[74] A. Ranfagni, D. Mugnai, *Phys, Lett. A* 322, 146 (2004).
[75] R. Mignani et al., *Rivista Nuovo Cimento* 4, 209–290, E398 1974).
[76] U. Bartocci, M.M. Capria, *American Journal of Physics* 59, 1030 (1991).
[77] S. Duro, F. Cardone, *Modern Physics Letters B* 28, 19, 1450156-1/8 (2014).
[78] G. Cherubini, A. Rosada and F. Cardone, *Journal of Advanced Physics* 7, 1–9 (2018).
[79] G. Albertini et al., *Journal of Radioanalytical and Nuclear Chemistry*, 304, 2, 955 (2015).
[80] F. Ridolfi, *Journal of Advanced Physics* 5,55 (2016).
[81] F. Cardone et al., *Int. J. Mod. Phys. B* 29, 1550239-1/13 (2015).
[82] F. Cardone et al., *Int. J. Mod. Phys. B* 13, 1750168-1/20 (2017).
[83] R. Capotosto, F. Rosetto, *Journal of Advanced Physics* 5, 80 (2016).
[84] G. Albertini, R. Capotosto, *Journal of Advanced Physics* 5, 84 (2016).
[85] A. Rosada, F. Cardone, *Modern Physics Letters B* 30, 28, 1650346-1/7 (2016).
[86] G. Albertini, D. Bassani, F. Cardone, *European Physical Journal Plus* 133, 39 (2018).
[87] F. Cardone, A. Carpinteri, G. Lacidogna, *Physics Letters A* 373, 4158–4163 (2009).
[88] F. Cardone et al., *Radiochimica Acta*, 107, 6, 469 (2019).
[89] A. Rosada et al., *Springer Nature Applied Science* 1, 1319 (2019).
[90] G. Albertini et al., *Int. J. Mod. Phys. B* 34, 04, 2050001 (2020).
[91] E. Storms, *Naturwissenschaften* 97, 861–881 (2010).
[92] D. Paci, C. Vedruccio, F. Ciciulla, D. Genovese, G. Albertini, Analysis and characterization of a twisted double loop antenna, arXiv:2002.08128.
[93] G. Albertini, et al., *Foundations of Science* (2020), https://doi.org/10.1007/s10699-020-09710-z.
[94] E. Komatsu et al., [WMAP Collaboration], Astrophys. J. Suppl. 192, 18 (2011);
P. A. R. Ade et al., Astron. Astrophys. 594, A13 (2014);
W. Perconti et al., Journal of Advanced Physics, 4, 1–5, (2015);
W. Perconti et al., Journal of Advanced Physics, 6, 1–5 (2016);
A. Petrucci et al., Journal of Advanced Physics, 7, 1–7 (2018).
[95] M. Gaspero, R. Mignani, F. Cardone, *European Physical Journal C* 4, 705–709 (1998).
[96] W. G. Scott Results on Jets from the UA1 Experiment. In: Cox B. (eds) QCD Hard Hadronic Processes. NATO ASI Series (Series B: Physics), vol 197. Springer, Boston, MA (1988);
M. Jonker, RAMPING '97, https://cds.cern.ch/record/382295/files/MJ2_07.PDF.
[97] F. Cardone, *Il Nuovo Cimento*, 104A, 757 (1991);

A. Di Domenico, Inverse Compton scattering of thermal radiation at LEP AND LEP-200, *Particle Accelerators* 39, 137-146 (1992).
[98] S. Bellucci, S. Ferrara, A. Shcherbakov, A. Yeranyan, *Physical Review D* 83, 065003 (2011).
[99] D. Colosi, C. Rovelli, *Classical and Quantum Gravity*, 26 (2), 025002 (2009).
[100] T. Padmanabhan, *Entropy*, 17, 7420-7452 (2015).